\documentclass[reprint,english,aps,apl,bibnotes]{revtex4-1}
\usepackage{graphicx}
\usepackage[T1]{fontenc}
\usepackage[latin9]{inputenc}
\usepackage{amstext}
\usepackage{amssymb}
\usepackage{multirow}
\usepackage{dcolumn}
\usepackage{color}
\usepackage{amsmath}
\usepackage{babel}
\makeatletter

\begin{document}

\title{Traveling wave parametric amplifier with Josephson junctions\\
       using minimal resonator phase matching}

\author{T.C. White$^1$}
\thanks{These authors contributed equally to this work}
\author{J.Y. Mutus$^{1,5}$}
\thanks{These authors contributed equally to this work}
\author{I.-C. Hoi$^1$}
\author{R. Barends$^{1,5}$}
\author{B. Campbell$^1$}
\author{Yu Chen$^{1,5}$}
\author{Z. Chen$^1$}
\author{B. Chiaro$^1$}
\author{A. Dunsworth$^1$}
\author{E. Jeffrey$^{1,5}$}
\author{J. Kelly$^1$}
\author{A. Megrant$^{1,2}$}
\author{C. Neill$^1$}
\author{P.J.J. O'Malley$^1$}
\author{P. Roushan$^{1,5}$}
\author{D. Sank$^{1,5}$}
\author{A. Vainsencher$^1$}
\author{J. Wenner$^1$}
\author{S. Chaudhuri$^{4}$}
\author{J. Gao$^{3}$}
\author{John M. Martinis$^{1,5}$}
\email{martinis@physics.ucsb.edu}
\affiliation{$^1$Department of Physics, University of California, Santa Barbara, CA 93106-9530}
\affiliation{$^2$Department of Materials, University of California, Santa Barbara, CA 93106}
\affiliation{$^3$National Institute of Standards and Technology, Boulder, CO 80305}
\affiliation{$^4$Department of Physics, Stanford University, Stanford, CA 94305}
\affiliation{$^5$Present address: Google, Santa Barbara, CA 93117}

\date{\today}
\begin{abstract}

Josephson parametric amplifiers have become a critical tool in superconducting device physics due to their high gain and quantum-limited noise. Traveling wave parametric amplifiers (TWPAs) promise similar noise performance while allowing for significant increases in both bandwidth and dynamic range. We present a TWPA device based on an LC-ladder transmission line of Josephson junctions and parallel plate capacitors using low-loss amorphous silicon dielectric. Crucially, we have inserted $\lambda/4$ resonators at regular intervals along the transmission line in order to maintain the phase matching condition between pump, signal, and idler and increase gain. We achieve an average gain of 12\,dB across a 4\,GHz span, along with an average saturation power of -92\,dBm with noise approaching the quantum limit.
  
\end{abstract}
\maketitle

The Josephson parametric amplifier \cite{yurke:originalJPA,yamamoto:fluxDrive,castellanos:JPA,hatridge:LJPA,abdo:jpc,JPADesign,eichler:dimer} (JPA) is a critical tool for high fidelity state measurement in superconducting qubits \cite{vijay:qJumps,hatridge:backaction,murch:trajectories} as it allows parametric amplification with near quantum-limited noise \cite{caves:qNoise}.  Despite its success, the JPA has typically been used only for single frequency measurements due to lower bandwidth and saturation power. A promising approach to scaling superconducting qubit experiments is frequency multiplexing \cite{day:broadband,chen:multiplexed,saira:entanglement}, which requires additional bandwidth and dynamic range for each measurement tone.   Simultaneous amplification of up to five multiplexed tones has been achieved with a JPA \cite{jeffrey:readout,barends:gates,9xMon} but was only possible with the Impedance-transformed parametric amplifier \cite{IMPA} (IMPA).  This highly engineered JPA provides much larger bandwidth and saturation power but pushes the resonant design to its low $Q$ limit.

To extend this frequency multiplexed approach for future experiments, we have adopted the distributed design of the traveling wave parametric amplifier (TWPA) \cite{yaakobi:JTWPA}.  Fiber-optic TWPAs have already demonstrated high gain, dynamic range, and bandwidth while reaching the quantum-limit of added noise \cite{Armstrong:fiberTWPA,kumar:opticalql}.  In this letter we present a microwave frequency TWPA with 4\,GHz of bandwidth and an order of magnitude more saturation power than the best JPA.  This device is compatible with scaling to much larger qubit systems through multiplexed measurement, and may find applications outside quantum information such as astrophysics detectors \cite{day:broadband,axion}

\begin{figure}
  
  \centering
    \includegraphics[width=0.48\textwidth]{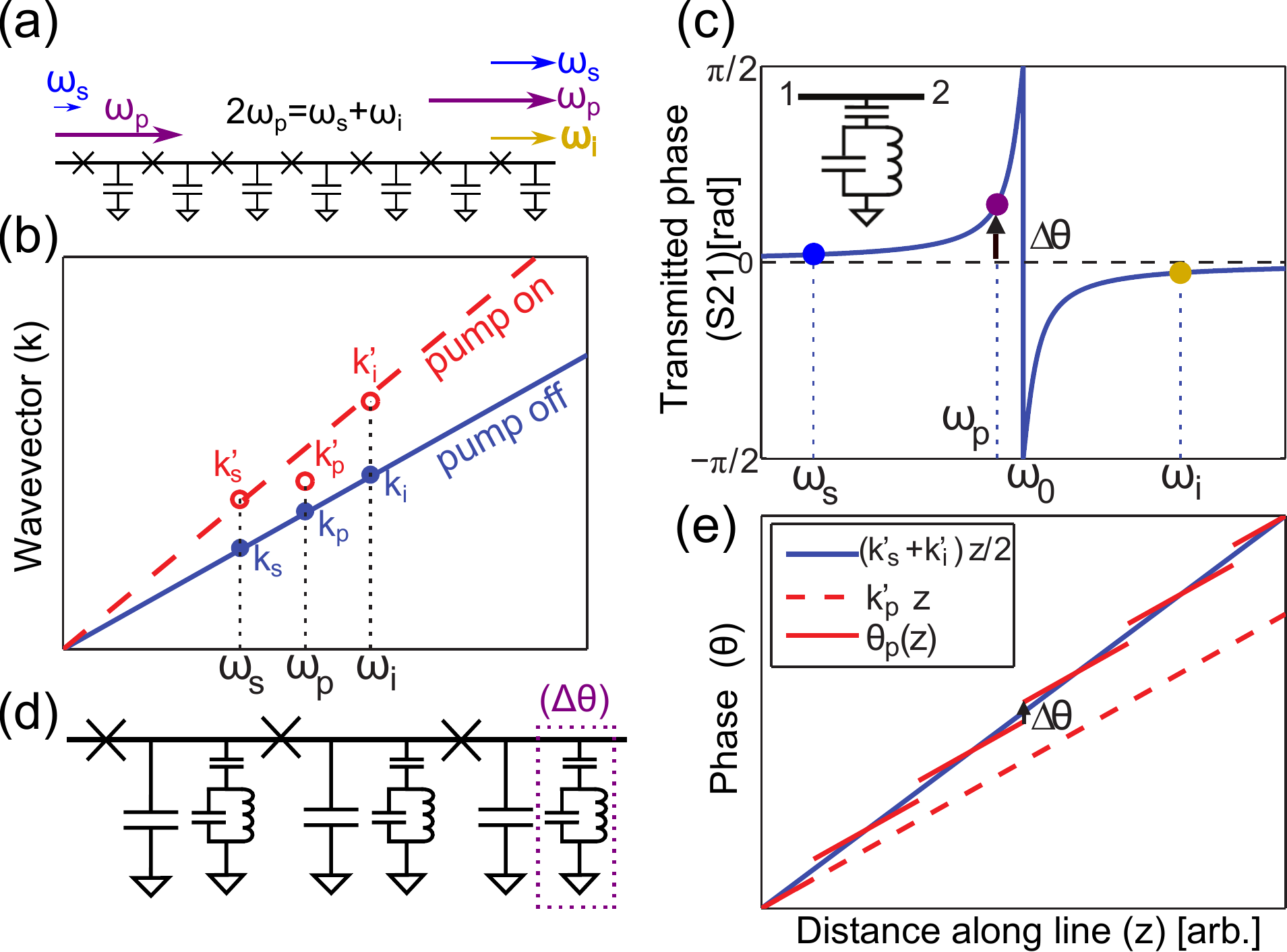}
  
\caption{(Color Online) (a)Circuit diagram of the Josephson juction TWPA. (b) The dispersion relationship in a conventional Josephson junction traveling wave parametric amplifier (TWPA). For weak signals (linear regime), the dispersion is linear $\Delta k = k_s + k_i - 2k_p = 0$ where $k = \omega/v_p$. When a strong pump tone is applied, the traveling waves are slowed down due to an increase in the junction inductance and a decrease in the phase velocity. The pump tone is slowed down less than the signal and idler tones due to the difference between the self-phase modulation and cross-phase modulation \cite{supp} effects which causes a mismatch $\Delta k' > 0$. (c) Phase shift due to a shunt resonator to ground. The resonator produces a frequency dependent phase shift for just the pump tone. (d) Resonantly phase matched TWPA in which resonator phase shifters are inserted between nonlinear transmission line sections. (e) The phase of the pump tone is adjusted at discrete locations and piece-wise matched to the signal and idler tones which enhances the gain, as can be seen from a tight fit between the stepped solid red line and the straight blue line $\Delta k = 0$.  Without these resonator phase shifters, the phase mismatch would grow (as can be seen from the departure of the dashed red line from the solid blue line) and the gain would be limited to the quadratic case of Eqn.\,(\ref{eqn:quad}).}
\label{fig:phaseMatching}
\end{figure}

At microwave frequencies the TWPA can be thought of as a transmission line where the propagation velocity is controlled by varying the individual circuit parameters of inductance or capacitance per unit length \cite{cullen:origTWPA,tien:orgTWPA2}. This is typically achieved by constructing a signal line with a current dependent (nonlinear) inductance.  Like the JPA, a large enough pump tone will modulate this inductance, coupling the pump ($\omega_p$) to a signal ($\omega_s$) and idler ($\omega_i$) tone via frequency mixing such that $\omega_s + \omega_i = 2\omega_p$.  Unlike the JPA however, the TWPA has no resonant structure so gain, bandwidth, and dynamic range are determined by the coupled mode equations of a nonlinear transmission line \cite{supp}.  In addition to allowing more bandwidth and saturation power, the TWPA is directional so that amplification only occurs for signals propagating in the same direction as the pump.

The concept of a nonlinear superconducting transmission line has been demonstrated in NbTiN TWPA \cite{eom:TiNparamp} where the kinetic inductance of the superconductor provides nonlinearity in a standard co-planar waveguide (CPW).  These amplifiers have achieved gains greater than 20\,dB over bandwidths greater than 8\,GHz, and with saturation power many orders of magnitude larger than a standard JPA.  To achieve this high dynamic range, a large pump tone ($\sim 100\,\mu W$) is required, which poses many engineering challenges for qubit readout. Attenuation in the line leads to heating which increases the base temperature of the experiment.  Likely due to this local heating, the NbTiN amplifier has yet to reach the quantum limit of added noise. In addition, the qubits must be aggressively isolated from the large pump tone which requires additional hardware.

\begin{figure*}
  
  \centering
    \includegraphics[width=\textwidth]{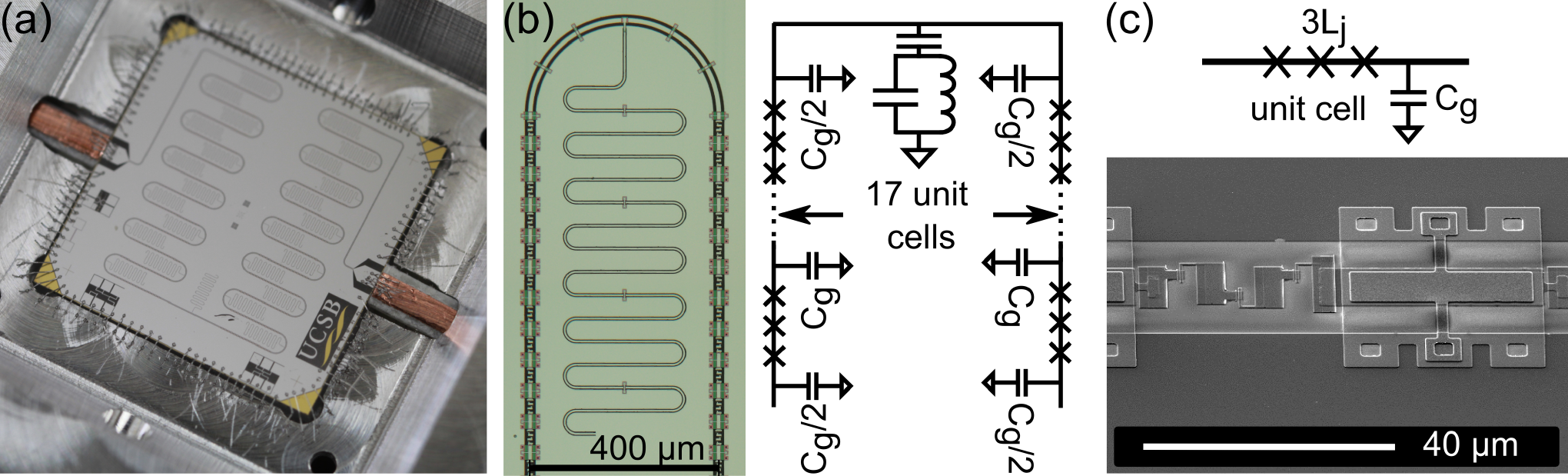}
\caption{(Color Online) (a) Photograph of the TWPA showing the full packaged device with an aluminum box and copper circuit-board feed lines.  The chip is square with 6\,mm sides.  (b) Optical micrograph and circuit diagram that show the discrete phase matching through the periodic insertion of $\lambda/4$ CPW resonators, spaced at an electrical length equivalent to $\lambda/2$ of the pump tone. (c) Scanning electron micrograph of the non-linear unit cell consisting of three double angle evaporation Josephson junctions(left) and a shunt parallel plate capacitor(right), with amorphous silicon dielectric.  }
  \label{fig:device}
\end{figure*}

The ideal amplifier for qubit readout would provide gain and bandwidth similar to the NbTiN TWPA but with a higher non-linearity, requiring less pump power and achieving quantum-limited noise.  A promising approach is to build a TWPA based on the non-linear inductance of the Josephson junction (JJ) \cite{wahlsten:expTWPA,feldman:juncTWPA,yurke:TWPA,yaakobi:TWPA,macklin:TWPA}. This junction TWPA (JTWPA) circuit, shown in Fig.\,\ref{fig:phaseMatching}(a), combines JJs with shunt capacitors to construct a 50\,$\Omega$ lumped element transmission line. The capacitors and JJs are small relative to the wavelength of a microwave signal, giving an effective capacitance and non-linear inductance per unit length.  The JTWPA obeys the same physics as the NbTiN TWPA, but needs $\sim 10^5$ times less pump power.

TWPA gain is described by solving the coupled mode equations including mixing terms between the pump($k_p$), signal($k_s$), and idler($k_i$) wave vectors \cite{supp}. Power gain is given by 
\begin{equation}
   \label{eqn:gain}
 G_s = \cosh^2(gz) + \left( \frac{\kappa}{2g} \right)^2\sinh^2(gz)
\end{equation}
and
\begin{equation}
	\label{eqn:lilg}
g = \sqrt{\frac{k_sk_i}{k_p^2}(\gamma k_p)^2 - (\kappa/2)^2}.
\end{equation}
Here $z$ is the length along the transmission line, $\gamma = I_p^2/16I_c^2$ describes the ratio of drive current to junction critical current (nonlinearity), and $\kappa = 2 \gamma k_p + k_s + k_i - 2k_p$ is the effective dispersion.  The pre-factor $k_sk_i/k_p^2$ describes the bandwidth of the amplifier and maximizes gain when $k_s = k_i = k_p$.  For small signal powers where $\gamma \approx 0$, $\kappa$ would be described by the difference in wave vectors $\Delta k = k_s+k_i - 2k_p$.  The term $2 \gamma k_p$ describes the self phase modulation of the pump, shown in Fig.\,\ref{fig:phaseMatching} (b), which increases with pump power.  In a linear superconducting transmission line there is no dispersion, so $\Delta k \approx 0$ making $\kappa \approx 2\gamma k_p$. The maximum gain then occurs when $g \approx 0$ and is given by 
\begin{equation}
 \label{eqn:quad}
G_s = 1 + (\gamma k_p z)^2 = 1+ \phi_{nl}^2,
\end{equation}
where $\phi_{nl} = \gamma k_p z$ is the nonlinear phase shift of the pump, such that gain depends quadratically on length.  If proper phase matching can be achieved, $\kappa = 0$, $g = \gamma k_p$, and the maximum gain, given by
\begin{equation}
  \label{eqn:exp}
  G_s = \cosh^2(\gamma k_p z) \approx \frac{\exp(2\phi_{nl})}{4},
\end{equation}
is exponentially dependent on length.  The phase matched design thus provides a much larger gain-bandwidth than a non phase matched TWPA for a given number of junctions and is a more efficient amplifier design.

To produce $\kappa = 0$ we can counter the power-dependent phase shift of the pump with an engineered frequency-tunable phase shift.  In the NbTiN TWPAs this was accomplished by a periodic impedance variation which created a narrow band gap and phase shift in the transmission \cite{bockstiege:tiNTWPA2}.  However, this approach provides only a small correction to the phase shift per unit length, which is incompatible with the high nonlinear phase shifts of the JTWPA.  Alternatively, a resonator capacitavly coupled to the transmission line, shown in Fig.\,\ref{fig:phaseMatching}(c), produces an arbitrarily large frequency dependent phase shift which counters the non-linear phase shift.  By including such a resonator after every nonlinear section in the transmission line, shown in Fig.\,\ref{fig:phaseMatching}(d), the pump frequency could be tuned to cancel the phase mismatch; thus making $\kappa \approx 0$. This approach has been shown to significantly increase both gain and bandwidth for a given number of junctions \cite{obrien:resTWPA, macklin:TWPA}.

While continuous phase correction is the most obvious approach, a resonator following each junction can introduce additional complications.  The large number of resonators would require a compact lumped-element design with parallel plate capacitors.  The frequency of these resonators would be harder to control and the extra dielectric will introduce more loss.  It should be possible however, to use fewer total resonators if we increase the phase shift from each individual resonator.  With fewer total resonators we can use larger CPW designs with lower loss and greater frequency control.  The concept of discrete phase correction is shown in Fig.\,\ref{fig:phaseMatching} (e).

Our device, shown in Fig.\,\ref{fig:device}(a), consists of a single 66\,mm CPW with both non-linear (lumped element JJ array) and linear (superconducting Al) sections.  The 1326 JJs are standard Al-$\text{Al}_2\text{O}_3$-Al junctions created using double angle evaporation \cite{dolan:bridge}.  The junction critical current was designed to be $5\, \mu \text{A}$ with an effective inductance of 65\,pH per junction that, combined with geometric inductance, gives 3.5\,$\mu \text{H/m}$. Parallel plate capacitors made with low loss amorphous silicon (a-Si:H) dielectric \cite{aaron:dielectric} provide 1.6\,$\text{nF/m}$ in the non-linear sections of the chip, setting the impedance while also shorting together the ground planes.  Connecting the ground planes is important because such a long transmission line could support lossy slot line modes.  The periodic structure shown in Fig.\,\ref{fig:device}(b) consists of a series of JJs followed by a linear section where the resonator is used to fix the pump phase before entering the next non-linear section.  To improve the impedance matching of the non-linear sections, the first and last shunt capacitor is half the capacitance of the others.  We chose a nonlinear unit cell of 3 junctions per capacitor, shown in Fig.\,\ref{fig:device}(c), to lower the transmission line cutoff frequency while maintaining a high junction critical current.  This was done to prevent leakage of pump power into higher harmonics, which may reduce gain, and to prevent the onset of a shock wave. \cite{landauer:shock}.

In the case of 1000 junctions, continuous phase matching can be approximated with the phase shift of just 10 ideal phase shifters \cite{supp}.  This result however does not consider the effect of the resonant amplitude dip on the pump, which can lead to large reflected pump energy.  Using more resonators will lessen this effect, but increase design complexity.  However, if the $\lambda/4$ resonators are spaced by $(2n)\lambda/4$, where $\lambda$ is the wavelength corresponding to the resonator frequency, the periodic placement provides a large stop band at 3$\omega$ to prevent pump leakage.  To take advantage of this enhancement on a chip with 1326 junctions, we chose 26 resonators with 17 non-linear unit cells between each resonator.  The resonance coming from the periodic placement combines with the resonators to create a sharp amplitude dip at 6.1\,GHz with optimal phase shift at about 5.8\,GHz, where almost no pump energy is reflected.

\begin{figure*}
  
  \centering
    \includegraphics[width=\textwidth]{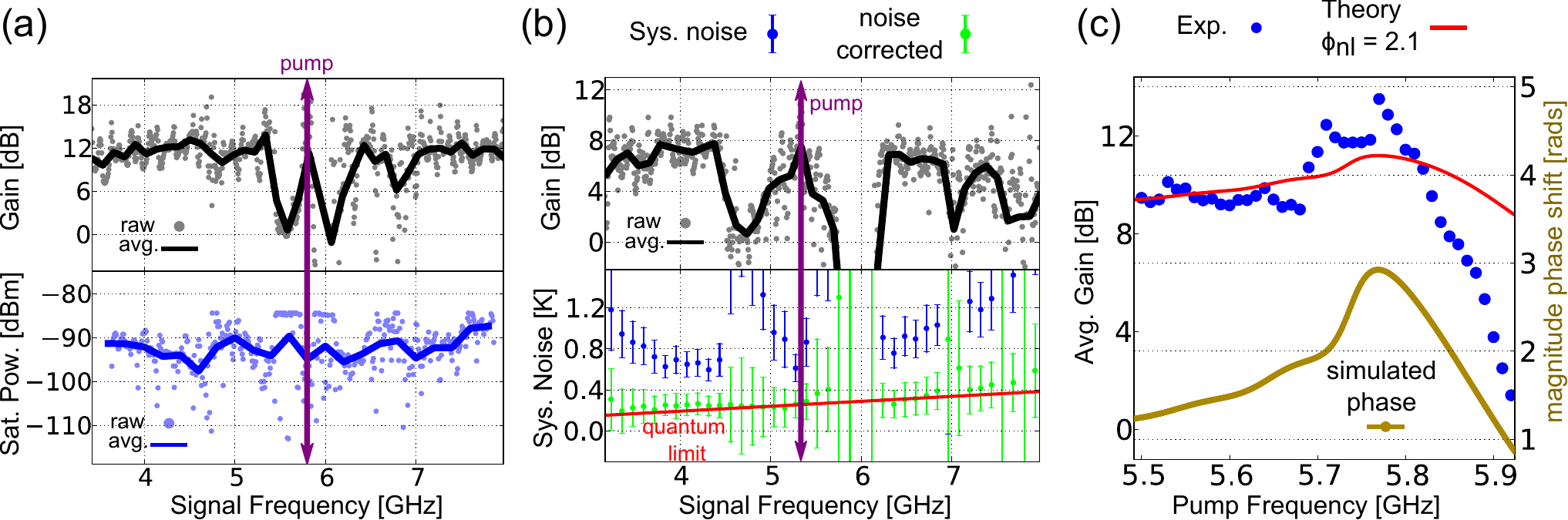}
\caption{(Color Online)(a) Saturation power and gain vs frequency for optimal pump gain with two TWPAs chained together; the pump tone is 5.83\,GHz.  Raw data is plotted in lighter points with a darker averaged line overlaying the data.  Average saturation power is -92\,dBm and average gain is 12-14\,dB.  The dips on either side of the pump come from the resonator reflecting either the signal or the idler tone close to the pump.  (b) Gain and system noise vs frequency for a different experiment optimized for low noise at a pump tone of 5.32\,GHz.  In this experiment the maximum gain achieved was 8\,dB while the lowest noise was 600\,mK corresponding to an input noise of 2 photons.  The raw system noise is plotted in blue(dark), while system noise with the contribution from the HEMT subtracted is plotted in green(light).  (c) Average gain and simulated phase shift vs frequency measured for a pair of devices which achieved 12-14\,dB max gain.  The blue (dark) data points are the averaged gain values, the yellow (light) curve is the simulated resonator phase shift, and the red (solid dark) line is a theoretical gain curve computed using the simulated phase shift.  The change from linear to exponential is consistent with $\phi_{nl}\approx 2.1$ with a peak at 5.8\,GHz.}
  \label{fig:performance}
\end{figure*}


In the past JTWPAs have had difficulty reaching the quantum limit of added noise due to loss in the transmission line \cite{yurke:TWPA}.  To characterize the loss and transmission line performance in our device, we measured the amplitude of $\text{S}_{21}$ and $\text{S}_{11}$ through both the TWPA and a copper cable of equivalent length.  We used \textit{in-situ} microwave switches to alternate between the cable and the TWPA in the same experimental setup \cite{supp}.  We find that the difference in $\text{S}_{21}$ is less than 0.5\,dB over the entire 4-8\,GHz measurement band.  When measuring $\text{S}_{11}$ of the TWPA we see an average reflected amplitude -10\,dB relative to the cable $\text{S}_{21}$.  This is consistent with the majority of the signal difference between the TWPA and the cable coming from reflections due to imperfect impedance matching \cite{supp}.

The device presented in Fig.\,\ref{fig:device}(a) provides a good test of amplifier performance, but is limited to 6-8\,dB gain. To increase the gain and verify the phase matching will hold in a longer device we chain two of these chips together in series.  The performance of this amplifier chain is shown in Fig.\,\ref{fig:performance} and details the 1\,dB compression point (saturation power), gain, and noise temperature.  Each chip was in a separate box and the boxes were connected via SMA connectors.  The gain was measured relative to the low power transmission amplitude.  

As can be seen in Fig\,\ref{fig:performance}(a), the chained device displays an average gain of 12-14\,dB over almost the entire 4-8\,GHz frequency range.  Interestingly the gain dips quite significantly on either side of the pump.  This is due to the reflection of either the signal or idler tone when measuring close to the pump frequency.  Variations in the gain on the order of 2-3\,dB most likely come from imperfect impedance matching between sections and at the bond pads.  These variations in gain also affect the saturation power, here defined as the 1\,dB compression point.  The broad band input saturation power shown in Fig.\,\ref{fig:performance}(a) varies from -95\,dBm to -85\,dBm with an average of -92 \,dBm.  This represents a significant improvement in both bandwidth and saturation power over the best resonant JPA \cite{IMPA}.  The reverse gain measured was 0\,dB as expected from the directionality of the coupled mode equations \cite{supp}.

To measure noise temperature we used the method of signal to noise ratio improvement \cite{hatridge:LJPA,supp} over a traditional high electron mobility (HEMT) semiconductor amplifier \cite{bradley:HEMT}.  In this experiment the HEMT noise temperature was measured to be $2.5\pm0.5$\,K over the measurement band. Unfortunately while measuring the TWPA noise temperature, the gain, shown in Fig.\,\ref{fig:performance} (b), reached only 8\,dB.  In this case we, find that the noise does approach the quantum limit over the entire range but reaches a low of only 600\,mK.  This noise temperature corresponds to about 2 photons of input noise and is consistent with residual HEMT noise at low gain.  If we subtract the expected HEMT contribution to the system noise we find the noise added by the TWPA is very close to quantum limit.

To verify the frequency dependence of the gain we measured maximum average gain vs frequency for the device shown in Fig.\,\ref{fig:performance}(a).  The frequency dependence of the gain is shown in Fig.\,\ref{fig:performance}(c) along with a simulated phase shift coming from the resonators.  The average gain increases by $\sim 5$\,dB when the pump nears the resonator phase shift.  This is consistent with a nonlinear phase shift $\phi_{nl} \approx 2.1$ with a predicted device gain plotted with a solid red line along with the data.

We have experimentally demonstrated a Josephson junction traveling wave parametric amplifier with minimal resonator phase matching.  This amplifier displays a significant increase in both bandwidth and saturation power while maintaining near quantum-limited noise performance.  By using discrete resonators to correct the pump phase we can access the exponential gain dependence with a minimal increase in fab complexity.  In this regime we should be able to increase the gain by simply increasing the length of the device.  In addition it may be possible to improve the transmission amplitude even further through fine tuning of the impedance in each section.


This work was supported by the Office of the Director of National Intelligence (ODNI), Intelligence Advanced Research Projects Activity (IARPA), through the Army Research Office grant W911NF-10-1-0334. All statements of fact, opinion or conclusions contained herein are those of the authors and should not be construed as representing the official views or policies of IARPA, the ODNI, or the U.S. Government.  Devices were made at the UC Santa Barbara Nanofabrication Facility, a part of the NSF-funded National Nanotechnology Infrastructure Network, and at the NanoStructures
Cleanroom Facility.


\begin{thebibliography}{37}
\expandafter\ifx\csname natexlab\endcsname\relax\def\natexlab#1{#1}\fi
\expandafter\ifx\csname bibnamefont\endcsname\relax
  \def\bibnamefont#1{#1}\fi
\expandafter\ifx\csname bibfnamefont\endcsname\relax
  \def\bibfnamefont#1{#1}\fi
\expandafter\ifx\csname citenamefont\endcsname\relax
  \def\citenamefont#1{#1}\fi
\expandafter\ifx\csname url\endcsname\relax
  \def\url#1{\texttt{#1}}\fi
\expandafter\ifx\csname urlprefix\endcsname\relax\def\urlprefix{URL }\fi
\providecommand{\bibinfo}[2]{#2}
\providecommand{\eprint}[2][]{\url{#2}}

\bibitem[{\citenamefont{Yurke et~al.}(1989)\citenamefont{Yurke, Corruccini,
  Kaminsky, Rupp, Smith, Silver, Simon, and Whittaker}}]{yurke:originalJPA}
\bibinfo{author}{\bibfnamefont{B.}~\bibnamefont{Yurke}},
  \bibinfo{author}{\bibfnamefont{L.}~\bibnamefont{Corruccini}},
  \bibinfo{author}{\bibfnamefont{P.}~\bibnamefont{Kaminsky}},
  \bibinfo{author}{\bibfnamefont{L.}~\bibnamefont{Rupp}},
  \bibinfo{author}{\bibfnamefont{A.}~\bibnamefont{Smith}},
  \bibinfo{author}{\bibfnamefont{A.}~\bibnamefont{Silver}},
  \bibinfo{author}{\bibfnamefont{R.}~\bibnamefont{Simon}}, \bibnamefont{and}
  \bibinfo{author}{\bibfnamefont{E.}~\bibnamefont{Whittaker}},
  \bibinfo{journal}{Phys. Rev. A} \textbf{\bibinfo{volume}{39}},
  \bibinfo{pages}{2519} (\bibinfo{year}{1989}).

\bibitem[{\citenamefont{Yamamoto et~al.}(2008)\citenamefont{Yamamoto, Inomata,
  Watanabe, Matsuba, Miyazaki, Oliver, Nakamura, and
  Tsai}}]{yamamoto:fluxDrive}
\bibinfo{author}{\bibfnamefont{T.}~\bibnamefont{Yamamoto}},
  \bibinfo{author}{\bibfnamefont{K.}~\bibnamefont{Inomata}},
  \bibinfo{author}{\bibfnamefont{M.}~\bibnamefont{Watanabe}},
  \bibinfo{author}{\bibfnamefont{K.}~\bibnamefont{Matsuba}},
  \bibinfo{author}{\bibfnamefont{T.}~\bibnamefont{Miyazaki}},
  \bibinfo{author}{\bibfnamefont{W.}~\bibnamefont{Oliver}},
  \bibinfo{author}{\bibfnamefont{Y.}~\bibnamefont{Nakamura}}, \bibnamefont{and}
  \bibinfo{author}{\bibfnamefont{J.}~\bibnamefont{Tsai}},
  \bibinfo{journal}{Appl. Phys. Lett.} \textbf{\bibinfo{volume}{93}},
  \bibinfo{pages}{042510} (\bibinfo{year}{2008}).

\bibitem[{\citenamefont{Castellanos-Beltran and
  Lehnert}(2007)}]{castellanos:JPA}
\bibinfo{author}{\bibfnamefont{M.}~\bibnamefont{Castellanos-Beltran}}
  \bibnamefont{and} \bibinfo{author}{\bibfnamefont{K.}~\bibnamefont{Lehnert}},
  \bibinfo{journal}{Appl. Phys. Lett.} \textbf{\bibinfo{volume}{91}},
  \bibinfo{pages}{083509} (\bibinfo{year}{2007}).

\bibitem[{\citenamefont{Hatridge et~al.}(2011)\citenamefont{Hatridge, Vijay,
  Slichter, Clarke, and Siddiqi}}]{hatridge:LJPA}
\bibinfo{author}{\bibfnamefont{M.}~\bibnamefont{Hatridge}},
  \bibinfo{author}{\bibfnamefont{R.}~\bibnamefont{Vijay}},
  \bibinfo{author}{\bibfnamefont{D.}~\bibnamefont{Slichter}},
  \bibinfo{author}{\bibfnamefont{J.}~\bibnamefont{Clarke}}, \bibnamefont{and}
  \bibinfo{author}{\bibfnamefont{I.}~\bibnamefont{Siddiqi}},
  \bibinfo{journal}{Phys. Rev. B} \textbf{\bibinfo{volume}{83}},
  \bibinfo{pages}{134501} (\bibinfo{year}{2011}).

\bibitem[{\citenamefont{Abdo et~al.}(2011)\citenamefont{Abdo, Schackert,
  Hatridge, Rigetti, and Devoret}}]{abdo:jpc}
\bibinfo{author}{\bibfnamefont{B.}~\bibnamefont{Abdo}},
  \bibinfo{author}{\bibfnamefont{F.}~\bibnamefont{Schackert}},
  \bibinfo{author}{\bibfnamefont{M.}~\bibnamefont{Hatridge}},
  \bibinfo{author}{\bibfnamefont{C.}~\bibnamefont{Rigetti}}, \bibnamefont{and}
  \bibinfo{author}{\bibfnamefont{M.}~\bibnamefont{Devoret}},
  \bibinfo{journal}{Appl. Phys. Lett.} \textbf{\bibinfo{volume}{99}},
  \bibinfo{pages}{162506} (\bibinfo{year}{2011}).

\bibitem[{\citenamefont{Mutus et~al.}(2013)\citenamefont{Mutus, White, Jeffrey,
  Sank, Barends, Bochmann, Chen, Chen, Chiaro, Dunsworth et~al.}}]{JPADesign}
\bibinfo{author}{\bibfnamefont{J.}~\bibnamefont{Mutus}},
  \bibinfo{author}{\bibfnamefont{T.}~\bibnamefont{White}},
  \bibinfo{author}{\bibfnamefont{E.}~\bibnamefont{Jeffrey}},
  \bibinfo{author}{\bibfnamefont{D.}~\bibnamefont{Sank}},
  \bibinfo{author}{\bibfnamefont{R.}~\bibnamefont{Barends}},
  \bibinfo{author}{\bibfnamefont{J.}~\bibnamefont{Bochmann}},
  \bibinfo{author}{\bibfnamefont{Y.}~\bibnamefont{Chen}},
  \bibinfo{author}{\bibfnamefont{Z.}~\bibnamefont{Chen}},
  \bibinfo{author}{\bibfnamefont{B.}~\bibnamefont{Chiaro}},
  \bibinfo{author}{\bibfnamefont{A.}~\bibnamefont{Dunsworth}},
  \bibnamefont{et~al.}, \bibinfo{journal}{Appl. Phys. Lett.}
  \textbf{\bibinfo{volume}{103}}, \bibinfo{pages}{122602}
  (\bibinfo{year}{2013}).

\bibitem[{\citenamefont{Eichler et~al.}(2014)\citenamefont{Eichler, Salathe,
  Mlynek, Schmidt, and Wallraff}}]{eichler:dimer}
\bibinfo{author}{\bibfnamefont{C.}~\bibnamefont{Eichler}},
  \bibinfo{author}{\bibfnamefont{Y.}~\bibnamefont{Salathe}},
  \bibinfo{author}{\bibfnamefont{J.}~\bibnamefont{Mlynek}},
  \bibinfo{author}{\bibfnamefont{S.}~\bibnamefont{Schmidt}}, \bibnamefont{and}
  \bibinfo{author}{\bibfnamefont{A.}~\bibnamefont{Wallraff}},
  \bibinfo{journal}{Phys. Rev. Lett.} \textbf{\bibinfo{volume}{113}},
  \bibinfo{pages}{110502} (\bibinfo{year}{2014}).

\bibitem[{\citenamefont{Vijay et~al.}(2011)\citenamefont{Vijay, Slichter, and
  Siddiqi}}]{vijay:qJumps}
\bibinfo{author}{\bibfnamefont{R.}~\bibnamefont{Vijay}},
  \bibinfo{author}{\bibfnamefont{D.}~\bibnamefont{Slichter}}, \bibnamefont{and}
  \bibinfo{author}{\bibfnamefont{I.}~\bibnamefont{Siddiqi}},
  \bibinfo{journal}{Phys. Rev. Lett.} \textbf{\bibinfo{volume}{106}},
  \bibinfo{pages}{110502} (\bibinfo{year}{2011}).

\bibitem[{\citenamefont{Hatridge et~al.}(2013)\citenamefont{Hatridge, Shankar,
  Mirrahimi, Schackert, Geerlings, Brecht, Sliwa, Abdo, Frunzio, Girvin
  et~al.}}]{hatridge:backaction}
\bibinfo{author}{\bibfnamefont{M.}~\bibnamefont{Hatridge}},
  \bibinfo{author}{\bibfnamefont{S.}~\bibnamefont{Shankar}},
  \bibinfo{author}{\bibfnamefont{M.}~\bibnamefont{Mirrahimi}},
  \bibinfo{author}{\bibfnamefont{F.}~\bibnamefont{Schackert}},
  \bibinfo{author}{\bibfnamefont{K.}~\bibnamefont{Geerlings}},
  \bibinfo{author}{\bibfnamefont{T.}~\bibnamefont{Brecht}},
  \bibinfo{author}{\bibfnamefont{K.}~\bibnamefont{Sliwa}},
  \bibinfo{author}{\bibfnamefont{B.}~\bibnamefont{Abdo}},
  \bibinfo{author}{\bibfnamefont{L.}~\bibnamefont{Frunzio}},
  \bibinfo{author}{\bibfnamefont{S.}~\bibnamefont{Girvin}},
  \bibnamefont{et~al.}, \bibinfo{journal}{Science}
  \textbf{\bibinfo{volume}{339}}, \bibinfo{pages}{178} (\bibinfo{year}{2013}).

\bibitem[{\citenamefont{Murch et~al.}(2013)\citenamefont{Murch, Weber, Macklin,
  and Siddiqi}}]{murch:trajectories}
\bibinfo{author}{\bibfnamefont{K.}~\bibnamefont{Murch}},
  \bibinfo{author}{\bibfnamefont{S.}~\bibnamefont{Weber}},
  \bibinfo{author}{\bibfnamefont{C.}~\bibnamefont{Macklin}}, \bibnamefont{and}
  \bibinfo{author}{\bibfnamefont{I.}~\bibnamefont{Siddiqi}},
  \bibinfo{journal}{Nature} \textbf{\bibinfo{volume}{502}},
  \bibinfo{pages}{211} (\bibinfo{year}{2013}).

\bibitem[{\citenamefont{Caves}(1982)}]{caves:qNoise}
\bibinfo{author}{\bibfnamefont{C.~M.} \bibnamefont{Caves}},
  \bibinfo{journal}{Phys. Rev. D} \textbf{\bibinfo{volume}{26}},
  \bibinfo{pages}{1817} (\bibinfo{year}{1982}).

\bibitem[{\citenamefont{Day et~al.}(2003)\citenamefont{Day, LeDuc, Mazin,
  Vayonakis, and Zmuidzinas}}]{day:broadband}
\bibinfo{author}{\bibfnamefont{P.~K.} \bibnamefont{Day}},
  \bibinfo{author}{\bibfnamefont{H.~G.} \bibnamefont{LeDuc}},
  \bibinfo{author}{\bibfnamefont{B.~A.} \bibnamefont{Mazin}},
  \bibinfo{author}{\bibfnamefont{A.}~\bibnamefont{Vayonakis}},
  \bibnamefont{and}
  \bibinfo{author}{\bibfnamefont{J.}~\bibnamefont{Zmuidzinas}},
  \bibinfo{journal}{Nature} \textbf{\bibinfo{volume}{425}},
  \bibinfo{pages}{817} (\bibinfo{year}{2003}).

\bibitem[{\citenamefont{Chen et~al.}(2012)\citenamefont{Chen, Sank, O'Malley,
  White, Barends, Chiaro, Kelly, Lucero, Mariantoni, Megrant
  et~al.}}]{chen:multiplexed}
\bibinfo{author}{\bibfnamefont{Y.}~\bibnamefont{Chen}},
  \bibinfo{author}{\bibfnamefont{D.}~\bibnamefont{Sank}},
  \bibinfo{author}{\bibfnamefont{P.}~\bibnamefont{O'Malley}},
  \bibinfo{author}{\bibfnamefont{T.}~\bibnamefont{White}},
  \bibinfo{author}{\bibfnamefont{R.}~\bibnamefont{Barends}},
  \bibinfo{author}{\bibfnamefont{B.}~\bibnamefont{Chiaro}},
  \bibinfo{author}{\bibfnamefont{J.}~\bibnamefont{Kelly}},
  \bibinfo{author}{\bibfnamefont{E.}~\bibnamefont{Lucero}},
  \bibinfo{author}{\bibfnamefont{M.}~\bibnamefont{Mariantoni}},
  \bibinfo{author}{\bibfnamefont{A.}~\bibnamefont{Megrant}},
  \bibnamefont{et~al.}, \bibinfo{journal}{Appl. Phys. Lett.}
  \textbf{\bibinfo{volume}{101}}, \bibinfo{pages}{182601}
  (\bibinfo{year}{2012}).

\bibitem[{\citenamefont{Saira et~al.}(2014)\citenamefont{Saira, Groen, Cramer,
  Meretska, De~Lange, and DiCarlo}}]{saira:entanglement}
\bibinfo{author}{\bibfnamefont{O.-P.} \bibnamefont{Saira}},
  \bibinfo{author}{\bibfnamefont{J.}~\bibnamefont{Groen}},
  \bibinfo{author}{\bibfnamefont{J.}~\bibnamefont{Cramer}},
  \bibinfo{author}{\bibfnamefont{M.}~\bibnamefont{Meretska}},
  \bibinfo{author}{\bibfnamefont{G.}~\bibnamefont{De~Lange}}, \bibnamefont{and}
  \bibinfo{author}{\bibfnamefont{L.}~\bibnamefont{DiCarlo}},
  \bibinfo{journal}{Phys. Rev. Lett.} \textbf{\bibinfo{volume}{112}},
  \bibinfo{pages}{070502} (\bibinfo{year}{2014}).

\bibitem[{\citenamefont{Jeffrey et~al.}(2014)\citenamefont{Jeffrey, Sank,
  Mutus, White, Kelly, Barends, Chen, Chen, Chiaro, Dunsworth
  et~al.}}]{jeffrey:readout}
\bibinfo{author}{\bibfnamefont{E.}~\bibnamefont{Jeffrey}},
  \bibinfo{author}{\bibfnamefont{D.}~\bibnamefont{Sank}},
  \bibinfo{author}{\bibfnamefont{J.}~\bibnamefont{Mutus}},
  \bibinfo{author}{\bibfnamefont{T.}~\bibnamefont{White}},
  \bibinfo{author}{\bibfnamefont{J.}~\bibnamefont{Kelly}},
  \bibinfo{author}{\bibfnamefont{R.}~\bibnamefont{Barends}},
  \bibinfo{author}{\bibfnamefont{Y.}~\bibnamefont{Chen}},
  \bibinfo{author}{\bibfnamefont{Z.}~\bibnamefont{Chen}},
  \bibinfo{author}{\bibfnamefont{B.}~\bibnamefont{Chiaro}},
  \bibinfo{author}{\bibfnamefont{A.}~\bibnamefont{Dunsworth}},
  \bibnamefont{et~al.}, \bibinfo{journal}{Phys. Rev. Lett.}
  \textbf{\bibinfo{volume}{112}}, \bibinfo{pages}{190504}
  (\bibinfo{year}{2014}).

\bibitem[{\citenamefont{Barends et~al.}(2014)\citenamefont{Barends, Kelly,
  Megrant, Veitia, Sank, Jeffrey, White, Mutus, Fowler, Campbell
  et~al.}}]{barends:gates}
\bibinfo{author}{\bibfnamefont{R.}~\bibnamefont{Barends}},
  \bibinfo{author}{\bibfnamefont{J.}~\bibnamefont{Kelly}},
  \bibinfo{author}{\bibfnamefont{A.}~\bibnamefont{Megrant}},
  \bibinfo{author}{\bibfnamefont{A.}~\bibnamefont{Veitia}},
  \bibinfo{author}{\bibfnamefont{D.}~\bibnamefont{Sank}},
  \bibinfo{author}{\bibfnamefont{E.}~\bibnamefont{Jeffrey}},
  \bibinfo{author}{\bibfnamefont{T.}~\bibnamefont{White}},
  \bibinfo{author}{\bibfnamefont{J.}~\bibnamefont{Mutus}},
  \bibinfo{author}{\bibfnamefont{A.}~\bibnamefont{Fowler}},
  \bibinfo{author}{\bibfnamefont{B.}~\bibnamefont{Campbell}},
  \bibnamefont{et~al.}, \bibinfo{journal}{Nature}
  \textbf{\bibinfo{volume}{508}}, \bibinfo{pages}{500} (\bibinfo{year}{2014}).

\bibitem[{\citenamefont{Kelly et~al.}(2015)\citenamefont{Kelly, Barends,
  Fowler, Megrant, Jeffrey, White, Sank, Mutus, Campbell, Chen et~al.}}]{9xMon}
\bibinfo{author}{\bibfnamefont{J.}~\bibnamefont{Kelly}},
  \bibinfo{author}{\bibfnamefont{R.}~\bibnamefont{Barends}},
  \bibinfo{author}{\bibfnamefont{A.}~\bibnamefont{Fowler}},
  \bibinfo{author}{\bibfnamefont{A.}~\bibnamefont{Megrant}},
  \bibinfo{author}{\bibfnamefont{E.}~\bibnamefont{Jeffrey}},
  \bibinfo{author}{\bibfnamefont{T.}~\bibnamefont{White}},
  \bibinfo{author}{\bibfnamefont{D.}~\bibnamefont{Sank}},
  \bibinfo{author}{\bibfnamefont{J.}~\bibnamefont{Mutus}},
  \bibinfo{author}{\bibfnamefont{B.}~\bibnamefont{Campbell}},
  \bibinfo{author}{\bibfnamefont{Y.}~\bibnamefont{Chen}}, \bibnamefont{et~al.},
  \bibinfo{journal}{Nature} \textbf{\bibinfo{volume}{519}}, \bibinfo{pages}{66}
  (\bibinfo{year}{2015}).

\bibitem[{\citenamefont{Mutus et~al.}(2014)\citenamefont{Mutus, White, Barends,
  Chen, Chen, Chiaro, Dunsworth, Jeffrey, and Kelly}}]{IMPA}
\bibinfo{author}{\bibfnamefont{J.}~\bibnamefont{Mutus}},
  \bibinfo{author}{\bibfnamefont{T.}~\bibnamefont{White}},
  \bibinfo{author}{\bibfnamefont{R.}~\bibnamefont{Barends}},
  \bibinfo{author}{\bibfnamefont{Y.}~\bibnamefont{Chen}},
  \bibinfo{author}{\bibfnamefont{Z.}~\bibnamefont{Chen}},
  \bibinfo{author}{\bibfnamefont{B.}~\bibnamefont{Chiaro}},
  \bibinfo{author}{\bibfnamefont{A.}~\bibnamefont{Dunsworth}},
  \bibinfo{author}{\bibfnamefont{E.}~\bibnamefont{Jeffrey}}, \bibnamefont{and}
  \bibinfo{author}{\bibfnamefont{J.~o.} \bibnamefont{Kelly}},
  \bibinfo{journal}{Appl. Phys. Lett.} \textbf{\bibinfo{volume}{104}},
  \bibinfo{pages}{263513} (\bibinfo{year}{2014}).

\bibitem[{\citenamefont{Yaakobi
  et~al.}(2013{\natexlab{a}})\citenamefont{Yaakobi, Friedland, Macklin, and
  Siddiqi}}]{yaakobi:JTWPA}
\bibinfo{author}{\bibfnamefont{O.}~\bibnamefont{Yaakobi}},
  \bibinfo{author}{\bibfnamefont{L.}~\bibnamefont{Friedland}},
  \bibinfo{author}{\bibfnamefont{C.}~\bibnamefont{Macklin}}, \bibnamefont{and}
  \bibinfo{author}{\bibfnamefont{I.}~\bibnamefont{Siddiqi}},
  \bibinfo{journal}{Phys. Rev. B} \textbf{\bibinfo{volume}{87}},
  \bibinfo{pages}{144301} (\bibinfo{year}{2013}{\natexlab{a}}).

\bibitem[{\citenamefont{Armstrong et~al.}(1962)\citenamefont{Armstrong,
  Bloembergen, Ducuing, and Pershan}}]{Armstrong:fiberTWPA}
\bibinfo{author}{\bibfnamefont{J.}~\bibnamefont{Armstrong}},
  \bibinfo{author}{\bibfnamefont{N.}~\bibnamefont{Bloembergen}},
  \bibinfo{author}{\bibfnamefont{J.}~\bibnamefont{Ducuing}}, \bibnamefont{and}
  \bibinfo{author}{\bibfnamefont{P.}~\bibnamefont{Pershan}},
  \bibinfo{journal}{Physical Review} \textbf{\bibinfo{volume}{127}},
  \bibinfo{pages}{1918} (\bibinfo{year}{1962}).

\bibitem[{\citenamefont{Kumar et~al.}(1990)\citenamefont{Kumar, Ayt\"ur, and
  Huang}}]{kumar:opticalql}
\bibinfo{author}{\bibfnamefont{P.}~\bibnamefont{Kumar}},
  \bibinfo{author}{\bibfnamefont{O.}~\bibnamefont{Ayt\"ur}}, \bibnamefont{and}
  \bibinfo{author}{\bibfnamefont{J.}~\bibnamefont{Huang}},
  \bibinfo{journal}{Phys. Rev. Lett.} \textbf{\bibinfo{volume}{64}},
  \bibinfo{pages}{1015} (\bibinfo{year}{1990}).

\bibitem[{\citenamefont{Asztalos et~al.}(2010)\citenamefont{Asztalos, Carosi,
  Hagmann, Kinion, Van~Bibber, Hotz, Rosenberg, Rybka, Hoskins, Hwang
  et~al.}}]{axion}
\bibinfo{author}{\bibfnamefont{S.~J.} \bibnamefont{Asztalos}},
  \bibinfo{author}{\bibfnamefont{G.}~\bibnamefont{Carosi}},
  \bibinfo{author}{\bibfnamefont{C.}~\bibnamefont{Hagmann}},
  \bibinfo{author}{\bibfnamefont{D.}~\bibnamefont{Kinion}},
  \bibinfo{author}{\bibfnamefont{K.}~\bibnamefont{Van~Bibber}},
  \bibinfo{author}{\bibfnamefont{M.}~\bibnamefont{Hotz}},
  \bibinfo{author}{\bibfnamefont{L.}~\bibnamefont{Rosenberg}},
  \bibinfo{author}{\bibfnamefont{G.}~\bibnamefont{Rybka}},
  \bibinfo{author}{\bibfnamefont{J.}~\bibnamefont{Hoskins}},
  \bibinfo{author}{\bibfnamefont{J.}~\bibnamefont{Hwang}},
  \bibnamefont{et~al.}, \bibinfo{journal}{Phys. Rev. Lett.}
  \textbf{\bibinfo{volume}{104}}, \bibinfo{pages}{041301}
  (\bibinfo{year}{2010}).

\bibitem[{sup()}]{supp}
\bibinfo{note}{See supplemental information at link for further description of
  sample, measurements, and equations}.

\bibitem[{\citenamefont{Cullen}(1958)}]{cullen:origTWPA}
\bibinfo{author}{\bibfnamefont{A.}~\bibnamefont{Cullen}}
  (\bibinfo{year}{1958}).

\bibitem[{\citenamefont{Tien}(1958)}]{tien:orgTWPA2}
\bibinfo{author}{\bibfnamefont{P.}~\bibnamefont{Tien}}, \bibinfo{journal}{J.
  Appl. Phys.} \textbf{\bibinfo{volume}{29}}, \bibinfo{pages}{1347}
  (\bibinfo{year}{1958}).

\bibitem[{\citenamefont{Eom et~al.}(2012)\citenamefont{Eom, Day, LeDuc, and
  Zmuidzinas}}]{eom:TiNparamp}
\bibinfo{author}{\bibfnamefont{B.~H.} \bibnamefont{Eom}},
  \bibinfo{author}{\bibfnamefont{P.~K.} \bibnamefont{Day}},
  \bibinfo{author}{\bibfnamefont{H.~G.} \bibnamefont{LeDuc}}, \bibnamefont{and}
  \bibinfo{author}{\bibfnamefont{J.}~\bibnamefont{Zmuidzinas}},
  \bibinfo{journal}{Nat. Phys.}  (\bibinfo{year}{2012}).

\bibitem[{\citenamefont{Wahlsten et~al.}(1977)\citenamefont{Wahlsten, Rudner,
  and Claeson}}]{wahlsten:expTWPA}
\bibinfo{author}{\bibfnamefont{S.}~\bibnamefont{Wahlsten}},
  \bibinfo{author}{\bibfnamefont{S.}~\bibnamefont{Rudner}}, \bibnamefont{and}
  \bibinfo{author}{\bibfnamefont{T.}~\bibnamefont{Claeson}},
  \bibinfo{journal}{Appl. Phys. Lett.} \textbf{\bibinfo{volume}{30}},
  \bibinfo{pages}{298} (\bibinfo{year}{1977}).

\bibitem[{\citenamefont{Feldman et~al.}(1975)\citenamefont{Feldman, Parrish,
  and Chiao}}]{feldman:juncTWPA}
\bibinfo{author}{\bibfnamefont{M.}~\bibnamefont{Feldman}},
  \bibinfo{author}{\bibfnamefont{P.}~\bibnamefont{Parrish}}, \bibnamefont{and}
  \bibinfo{author}{\bibfnamefont{R.}~\bibnamefont{Chiao}}, \bibinfo{journal}{J.
  Appl. Phys.} \textbf{\bibinfo{volume}{46}}, \bibinfo{pages}{4031}
  (\bibinfo{year}{1975}).

\bibitem[{\citenamefont{Yurke et~al.}(1996)\citenamefont{Yurke, Roukes,
  Movshovich, and Pargellis}}]{yurke:TWPA}
\bibinfo{author}{\bibfnamefont{B.}~\bibnamefont{Yurke}},
  \bibinfo{author}{\bibfnamefont{M.}~\bibnamefont{Roukes}},
  \bibinfo{author}{\bibfnamefont{R.}~\bibnamefont{Movshovich}},
  \bibnamefont{and}
  \bibinfo{author}{\bibfnamefont{A.}~\bibnamefont{Pargellis}},
  \bibinfo{journal}{Appl. Phys. Lett.} \textbf{\bibinfo{volume}{69}},
  \bibinfo{pages}{3078} (\bibinfo{year}{1996}).

\bibitem[{\citenamefont{Yaakobi
  et~al.}(2013{\natexlab{b}})\citenamefont{Yaakobi, Friedland, Macklin, and
  Siddiqi}}]{yaakobi:TWPA}
\bibinfo{author}{\bibfnamefont{O.}~\bibnamefont{Yaakobi}},
  \bibinfo{author}{\bibfnamefont{L.}~\bibnamefont{Friedland}},
  \bibinfo{author}{\bibfnamefont{C.}~\bibnamefont{Macklin}}, \bibnamefont{and}
  \bibinfo{author}{\bibfnamefont{I.}~\bibnamefont{Siddiqi}},
  \bibinfo{journal}{Phys. Rev. B} \textbf{\bibinfo{volume}{87}},
  \bibinfo{pages}{144301} (\bibinfo{year}{2013}{\natexlab{b}}).

\bibitem[{\citenamefont{Macklin and et~al.}(2014)}]{macklin:TWPA}
\bibinfo{author}{\bibfnamefont{C.}~\bibnamefont{Macklin}} \bibnamefont{and}
  \bibinfo{author}{\bibnamefont{et~al.}}, \bibinfo{journal}{in preparation}
  (\bibinfo{year}{2014}).

\bibitem[{\citenamefont{Bockstiegel et~al.}(2014)\citenamefont{Bockstiegel,
  Gao, Vissers, Sandberg, Chaudhuri, Sanders, Vale, Irwin, and
  Pappas}}]{bockstiege:tiNTWPA2}
\bibinfo{author}{\bibfnamefont{C.}~\bibnamefont{Bockstiegel}},
  \bibinfo{author}{\bibfnamefont{J.}~\bibnamefont{Gao}},
  \bibinfo{author}{\bibfnamefont{M.}~\bibnamefont{Vissers}},
  \bibinfo{author}{\bibfnamefont{M.}~\bibnamefont{Sandberg}},
  \bibinfo{author}{\bibfnamefont{S.}~\bibnamefont{Chaudhuri}},
  \bibinfo{author}{\bibfnamefont{A.}~\bibnamefont{Sanders}},
  \bibinfo{author}{\bibfnamefont{L.}~\bibnamefont{Vale}},
  \bibinfo{author}{\bibfnamefont{K.}~\bibnamefont{Irwin}}, \bibnamefont{and}
  \bibinfo{author}{\bibfnamefont{D.}~\bibnamefont{Pappas}},
  \bibinfo{journal}{J. Low. Temp. Phys.} pp. \bibinfo{pages}{1--7}
  (\bibinfo{year}{2014}).

\bibitem[{\citenamefont{O'Brien et~al.}(2014)\citenamefont{O'Brien, Macklin,
  Siddiqi, and Zhang}}]{obrien:resTWPA}
\bibinfo{author}{\bibfnamefont{K.}~\bibnamefont{O'Brien}},
  \bibinfo{author}{\bibfnamefont{C.}~\bibnamefont{Macklin}},
  \bibinfo{author}{\bibfnamefont{I.}~\bibnamefont{Siddiqi}}, \bibnamefont{and}
  \bibinfo{author}{\bibfnamefont{X.}~\bibnamefont{Zhang}},
  \bibinfo{journal}{arXiv preprint arXiv:1406.2346}  (\bibinfo{year}{2014}).

\bibitem[{\citenamefont{Dolan}(1977)}]{dolan:bridge}
\bibinfo{author}{\bibfnamefont{G.}~\bibnamefont{Dolan}},
  \bibinfo{journal}{Appl. Phys. Lett.} \textbf{\bibinfo{volume}{31}},
  \bibinfo{pages}{337} (\bibinfo{year}{1977}).

\bibitem[{\citenamefont{O'Connell et~al.}(2008)\citenamefont{O'Connell,
  Ansmann, Bialczak, Hofheinz, Katz, Lucero, McKenney, Neeley, Wang, Weig
  et~al.}}]{aaron:dielectric}
\bibinfo{author}{\bibfnamefont{A.~D.} \bibnamefont{O'Connell}},
  \bibinfo{author}{\bibfnamefont{M.}~\bibnamefont{Ansmann}},
  \bibinfo{author}{\bibfnamefont{R.~C.} \bibnamefont{Bialczak}},
  \bibinfo{author}{\bibfnamefont{M.}~\bibnamefont{Hofheinz}},
  \bibinfo{author}{\bibfnamefont{N.}~\bibnamefont{Katz}},
  \bibinfo{author}{\bibfnamefont{E.}~\bibnamefont{Lucero}},
  \bibinfo{author}{\bibfnamefont{C.}~\bibnamefont{McKenney}},
  \bibinfo{author}{\bibfnamefont{M.}~\bibnamefont{Neeley}},
  \bibinfo{author}{\bibfnamefont{H.}~\bibnamefont{Wang}},
  \bibinfo{author}{\bibfnamefont{E.~M.} \bibnamefont{Weig}},
  \bibnamefont{et~al.}, \bibinfo{journal}{Appl. Phys. Lett.}
  \textbf{\bibinfo{volume}{92}}, \bibinfo{pages}{112903}
  (\bibinfo{year}{2008}).

\bibitem[{\citenamefont{Landauer}(1960)}]{landauer:shock}
\bibinfo{author}{\bibfnamefont{R.}~\bibnamefont{Landauer}},
  \bibinfo{journal}{IBM Journal of Research and Development}
  \textbf{\bibinfo{volume}{4}}, \bibinfo{pages}{391} (\bibinfo{year}{1960}).

\bibitem[{\citenamefont{Bradley}(1999)}]{bradley:HEMT}
\bibinfo{author}{\bibfnamefont{R.}~\bibnamefont{Bradley}},
  \bibinfo{journal}{Nuclear Phys. B-Proceedings Supplements}
  \textbf{\bibinfo{volume}{72}}, \bibinfo{pages}{137} (\bibinfo{year}{1999}).

\end{thebibliography}
\end{document}


\title{Supplementary Information}

\date{\today}

\section{Supplementary Information}
\subsection{Coupled Mode Equations}
A circuit diagram of a Josephson junction embedded nonlinear transmission line is shown in Fig.~1(a) in the main text. The Josephson inductance is current dependent,
\begin{equation}\label{eqn:nlininductance}
L(I) = L _{0} \left[1 + \frac{1}{2} \frac{I^{2}}{I_{c}^{2}} \right], L_{0}=\frac{\Phi_{0}}{2\pi I_{c}}
\end{equation}
where $\Phi_{0}$ is the magnetic flux quantum and $I_{c}$ is the critical current of the Josephson junction. Consequently, propagation along the transmission line is described by a nonlinear wave equation:
\begin{eqnarray}
\frac{\partial^2 I}{\partial z^2}-\frac{1}{\tilde{c}^2}\frac{\partial^2 }{\partial t^2}\left[ I + \frac{1}{6} \frac{I^3}{I_{c}^2}\right] = 0,~\frac{1}{\tilde{c}^2}\approx LC,\label{eqn:nlinwaveeqn}
\end{eqnarray}
Eqn.~\ref{eqn:nlinwaveeqn} is analogous to the case of light traveling in nonlinear Kerr media, in which the index of refraction is intensity-dependent. The propagation can be solved using the coupled-mode equations (CME) method from nonlinear optics. \cite{agrawal}

We write the pump, signal and idler as $I = 1/2\left[\sum_n A_n e^{i(k_n z - \omega_n t)} + c.c.\right]$ where $A_p$, $A_s$ and $A_i$ represent the three traveling waves of pump, signal and idler, respectively. Following the CME approach, we derive the coupled-mode equations under the slow wave approximation (SWA), \cite{eom:TiNparamp} \cite{cgi}
\begin{eqnarray}
\frac{d A_p}{d z} &=& \frac{i k_p}{16 I_{c}^2 } A_p|A_p|^2\nonumber\\
\frac{d A_s}{d z} &=& \frac{i k_s}{16 I_{c}^2 }\left(2A_s|A_p|^2+A_i^*A_p^2 e^{-i\Delta k z}\right)\nonumber\\
\frac{d A_i}{d z} &=& \frac{i k_i}{16 I_{c}^2 }\left(2A_i|A_p|^2+A_s^*A_p^2e^{-i\Delta k z}\right),
\end{eqnarray}
where $\Delta k = k_i + k_s -2k_p$ is the phase mismatch calculated from weak-signal dispersion of the transmission line. The term on the right hand side of the pump equation represents the self-phase modulation due to the AC pump current interacting with itself. Analogous cross-phase modulation processes are represented by the first term on the right hand side of the signal and idler equations; the other term presents the conversion of two pump photons to a signal photon and an idler photon. Under the undepleted pump assumption $|A_p|\gg |A_i|,|A_s|$, the pump can be solved first
\begin{eqnarray}
A_p = A_p(0)e^{i\gamma k_p z},\gamma = \frac{|A_p(0)|^2}{16 I_{c}^2},
\end{eqnarray}
where $\gamma$ is an unit-less coefficient reflecting the strength of the nonlinear effect and $\phi_{nl} = \gamma k_p z$ is the nonlinear phase shift of the pump also referred to as the self phase modulation (SPM) in nonlinear optics. Using the pump solution, the signal and idler can be solved perturbatively. Here, we write the solution in a matrix form,
\begin{eqnarray}
\left[
\begin{matrix}
A_s(z)
\\
A_i^*(z)
\end{matrix}
\right]
&=& M(z, A_p(0))
\left[
\begin{matrix}
A_s(0)
\\
A_i^*(0)
\end{matrix}
\right],~
M
=
\left[
\begin{matrix}
u_{11} & u_{12}
\\
u_{21} &  u_{22}
\end{matrix}
\right]\nonumber\\
u_{11} &=& \left[\cosh(g z) + i \frac{\kappa}{2g}\sinh(g z)\right]e^{i(2\gamma k_s-\frac{\kappa}{2})z},\nonumber\\
u_{12} &=& \left[\frac{i\gamma k_s}{g}e^{2i\phi_0}\sinh(gz)\right]e^{i(2\gamma k_s-\frac{\kappa}{2})z},\nonumber\\
u_{21} &=& \left[-\frac{i\gamma k_i}{g}e^{-2i\phi_0}\sinh(gz)\right]e^{-i(2\gamma k_i-\frac{\kappa}{2})z},\nonumber\\
u_{22} &=& \left[\cosh(g z) - i \frac{\kappa}{2g}\sinh(g z)\right]e^{-i(2\gamma k_i-\frac{\kappa}{2})z},\nonumber\\
e^{i\phi_0} &=& \frac{A_{p0}}{|A_{p0}|},~g = \sqrt{\left( \frac{k_s k_i}{k_p^2} \right)(\gamma k_p)^2- \left(\frac{\kappa}{2}\right)^2}\label{eqn:uvgeneral},~\kappa = 2\gamma k_p + \Delta k \label{eqn:kappageneral}.
\end{eqnarray}

If the transmission line has no intrinsic dispersion ($\Delta k = 0$, as is our case for the junction embedded transmission line in the low frequency limit), it can be derived from Eqn.\,\ref{eqn:uvgeneral} that the maximum signal gain (occurring at $\omega_{s}=\omega_{p}$) is quadratic in $\phi_{nl}$ or $z$ (thus the length of the line)\cite{yaakobi:JTWPA, chaudhurigao} ,
\begin{eqnarray}
g & \rightarrow & 0,~u_{11}= (1 + i \gamma k_p z) e^{i\gamma k_p z},~G_q = |u_{11}|^2 = 1 + \phi^2_{nl}.
\end{eqnarray}

If additional dispersion $\Delta k = -2\gamma k_p$ is introduced (as is the case in the dispersion engineered kinetic inductance parametric amplifier \cite{eom:TiNparamp}) so that the phase matching condition is perfectly met ($\kappa = 0$), exponential signal gain can be achieved,
\begin{eqnarray}
G_{e} &=& \cosh^2(\phi_{nl}) \approx \exp(2\phi_{nl})/4.
\end{eqnarray}

\subsection{Resonantly Phase-matched TWPA}
\begin{figure}
\centering
\includegraphics[width=\textwidth]{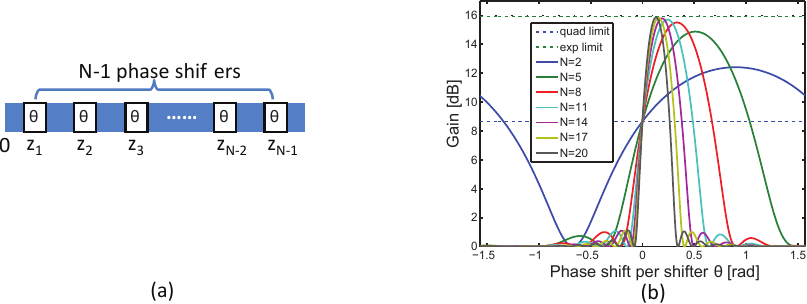}
\caption{(a) Configuration of $N-1$ phase shifters inserted between $N$ nonlinear transmission line sections. (b)  Comparison of enhanced gain with the quadratic gain and exponential gain limits. $\phi_{nl} = 2.5$ is assumed which corresponds to our device.}
\label{fig:phaseshifter}.
\end{figure}

Assume that N - 1 phase shifters (resonators) are inserted in between N sections of dispersionless line at position $z_m$ (see Figure~{\ref{fig:phaseshifter}}). We treat the resonators as perfect phase shifters for which $S_{21} = 1$ at all frequencies except the pump, where we have $S_{21}(\omega_p) = e^{i\theta}$. The output signal/idler at $z_N$ and the total signal gain can be calculated by cascading the M matrices
\begin{eqnarray}
\left[
\begin{matrix}
A_s(z_N)
\\
A_i^*(z_N)
\end{matrix}
\right]
&=& M(z_N-z_{N-1},~A_p(z^+_{N-1}))M(z_{N-1}-z_{N-2},~A_p(z^+_{N-2}))...M(z_1,~A_p(0))
\left[
\begin{matrix}
A_s(0)
\\
A_i^*(0)
\end{matrix}
\right]\nonumber\\
&=& \hat{M}(z_N, A_p(0))\left[
\begin{matrix}
A_s(0)
\\
A_i^*(0)
\end{matrix}
\right],~G_{r} = |\hat{M}_{11}|^2.
\end{eqnarray}
where $A_p(z^+_m)$ has included the additional phase shift $\theta$ from the phase shifter at $z_m$ and $\hat{M}$ is the cumulative transfer matrix for signal/idler from $z=0$ to $z=z_N$. A Matlab program is written to compute $\hat{M}$ and $G_{r}$. We are mostly interested in the dependence of $G_{r}$ on the number of phase shifters $N$ and the phase shift per shifter $\theta$.
\begin{figure}
\centering
\includegraphics[width=\textwidth]{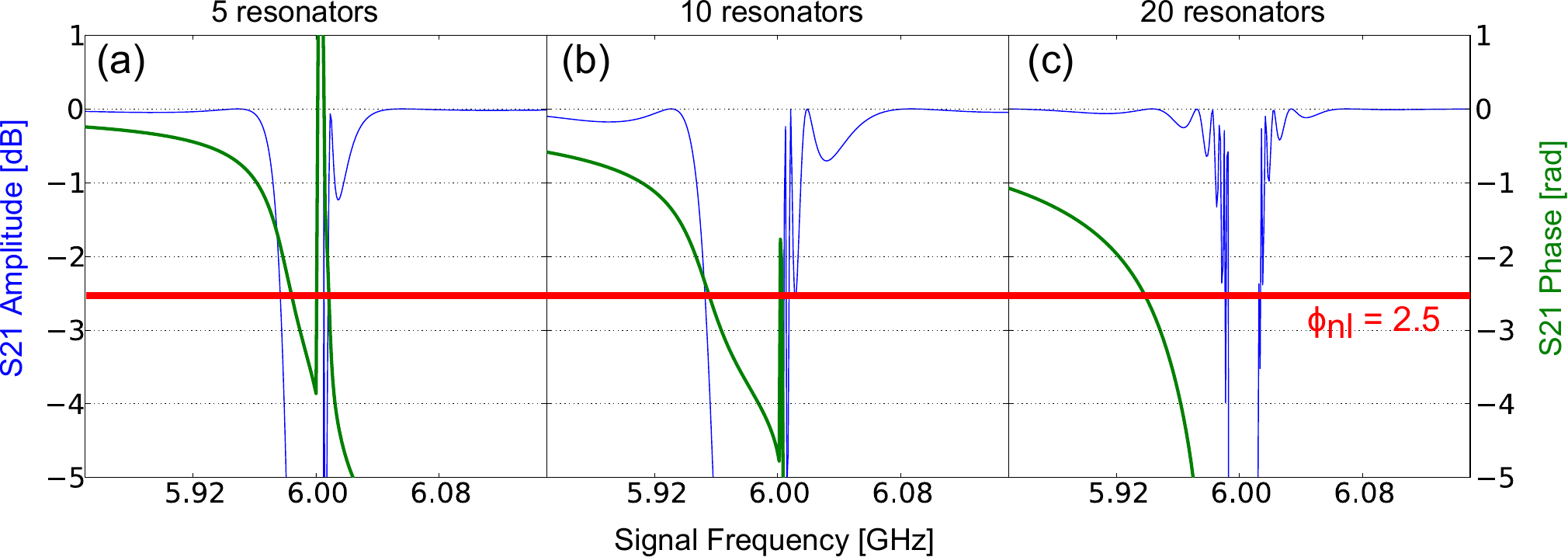}
\caption{Plots showing simulated signal phase and amplitude vs frequency for (a) 5 resonators, (b) 10 resonators, or (c) 20 resonators in a TWPA circuit.  The plots show that while only a few ideal phase shifters are necessary for phase matching,  $\lambda/4$ resonators also cause an amplitude dip which can affect pump transmission close to resonance.  Adding additional resonators allows for a greater phase shift with a smaller reflected pump amplitude.  The red line indicates a nonlinear phase shift of 2.5 radians which is necessary to achieve gain greater than 15 dB.}
\label{fig:resNumber}.
\end{figure}
Fig.\,\ref{fig:phaseshifter} (b) shows the calculation results, using the realistic design parameters of the device. It is clear that the gain is greatly enhanced by the phase shifters, even when only 1 phase shifter ($N=2$) is inserted. The enhancement increases with $N$ and approaches the exponential gain limit for large $N$. In fact, the enhanced gain is very close to $G_e$ limit (green dashed line) for 7 or more phase shifters ($N>8$). Adding resonators per unit LC ladder was recently proposed in \cite{obrien:resTWPA} to achieve phase matching condition.

An ideal phase shifter can be approximated by a $\lambda/4$ resonator capacitively coupled to the transmission line.  The resonator provides a frequency dependent phase shift as well as an amplitude dip which is maximized on resonance.  With only a few resonators 5-10, it is impossible to achieve the desired phase shift without tuning the pump into the amplitude dip of the resonator.  This will destroy the parametric gain through internal reflections in the transmission line.  To ensure we can achieve the desired phase shift with virtually no affect on pump amplitude we must use a design with $N\geq 20$.  The dependence of phase and amplitude on frequency is shown for several numbers of resonators in Fig.\,\ref{fig:resNumber}.

\subsection{TWPA Device Parameters}

Constructing the nonlinear sections of the TWPA requires balancing between the critical current of the Josephson junctions and the cutoff frequency of the LC ladder. If the critical current is increased it will require a larger pump to achieve the same nonlinearity.  Saturation power depends directly on pump amplitude so a higher critical current for each junction is desirable.  However higher critical current also means a lower inductance per section in the LC ladder which means a higher cutoff frequency given by $1/(2 \pi \sqrt{L_{sec} C_{sec}})$. A lower cutoff frequency is desirable because it will prevent parasitic coupling of the pump to higher frequency modes\cite{landauer:shock}.  Thus we constructed each section with three higher critical current junctions in a row, such that $L_{sec} = 3L_j$.  This allowed us to use higher critical current junction while also lowering the cutoff frequency of the transmission line by a factor of three.  The capacitance was then increased to maintain $\sqrt{\mathcal{L}/\mathcal{C}} \approx 50\,\Omega$ given inductance per unit length $\mathcal{L}$ and capacitance per unit length $\mathcal{C}$

As stated in the main text the critical current of each junction was designed to be $\approx 5\, \mu \text{A}$ which corresponds to an inductance of 65\,pH and a section inductance of 195\,pH.  The capacitance of each parallel plate capacitor was designed to be 117\, fF leading to a cutoff frequency of 33\,GHz.  This cutoff frequency combined with the dispersion engineering is sufficient to prevent the propagation of shock-waves in the transmission line\cite{landauer:shock}.  The geometric inductance and capacitance per unit length were extracted from simulations matching the nonlinear transmission line geometry.  Combining the simulation data with the single section values give $\mathcal{L} = 3.5\, \mu \text{H/m}$ and $\mathcal{C} = 1.5\,\text{nF/m}$ for a combined impedance of $\approx 48\, \Omega$.  The impedance was designed to be less than $50\, \Omega$ initially as impedance will increase slightly when in operation do to the nonlinear inductance.

The resonators were initially designed to operate at a frequency of 7\,GHz but shifted lower in frequency to 6.1\,GHz due to kinetic inductance in the thin 60\,nm aluminum film. Resonators being placed at the end of each nonlinear section means they are $1100\,\mu \text{m}$ apart.   The propagation velocity coming from the inductance and capacitance per unit length means $1100\,\mu \text{m}$ corresponds to $\lambda/2$ for 6.2\,GHz.  This is consistent with what we observe experimentally.

\subsection{Measuring TWPA Transmission}

\begin{figure}
\centering
\includegraphics[width=0.5\textwidth]{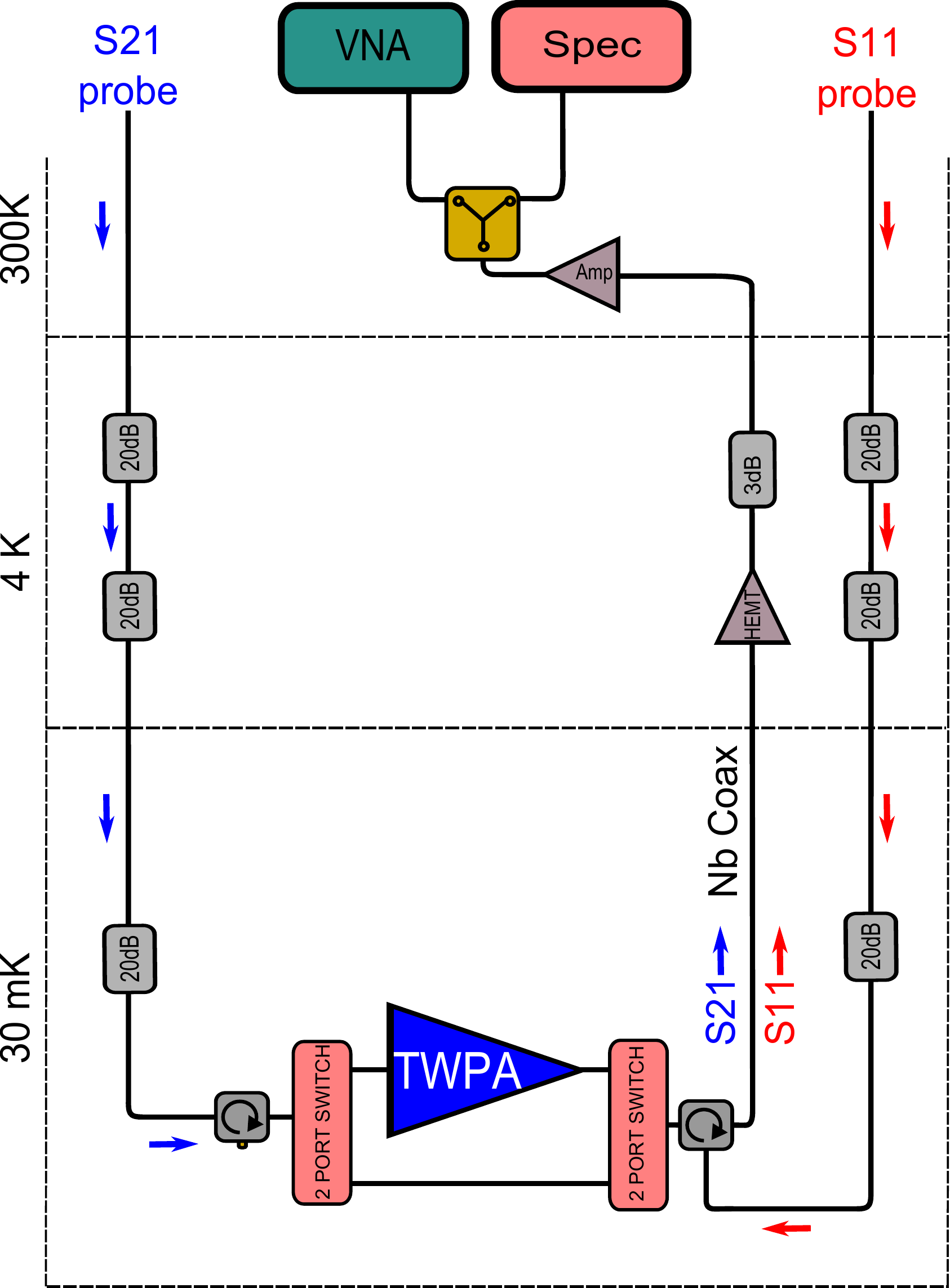}
\caption{Diagram of the transmission measurement experiment used to characterize TWPA transmission and reflection amplitudes.  The output port of a vector network analyzer (VNA) can be attached to either port and combined with the TWPA pump.  The signal output is then split at room temperature between the VNA and a spectrum analyzer (SPEC) used for noise measurements.  This setup allows us to probe both $\text{S}_{21}$ and $\text{S}_{11}$ in a single cool down of the refrigerator using equivalent measurement paths. }
\label{fig:switches}
\end{figure}

For the TWPA to function as an effective amplifier it must first function as a transmission line.  If the individual sections are not well impedance matched, internal reflections can destroy the coupling between different frequency modes.  Excessive loss in the line can become a source of noise which will make quantum-limited amplification impossible.  To check the transmission line behavior of the TWPA we measured its transmission ($\text{S}_{21}$) and reflection ($\text{S}_{11}$) amplitude compared to that of a standard low loss microwave cable.  This experiment was carried out using two 2-port cryogenic microwave switches as well as two cryogenic microwave circulators, shown in Fig.\,\ref{fig:switches}.  The switches were used to swap the TWPA and cable in the transmission path to the high electron mobility transistor (HEMT) amplifier.  Two equivalent microwave inputs were used to probe the transmission and reflection simultaneously.  The $\text{S}_{21}$ probe line goes through the first circulator, the TWPA, and finally the last circulator, before going to the HEMT.  The $\text{S}_{11}$ line goes to the third input of the last circulator which funnels it to the opposite end of the TWPA. Any reflected signal then makes its way to the HEMT.

\begin{figure}
\centering
\includegraphics[width=\textwidth]{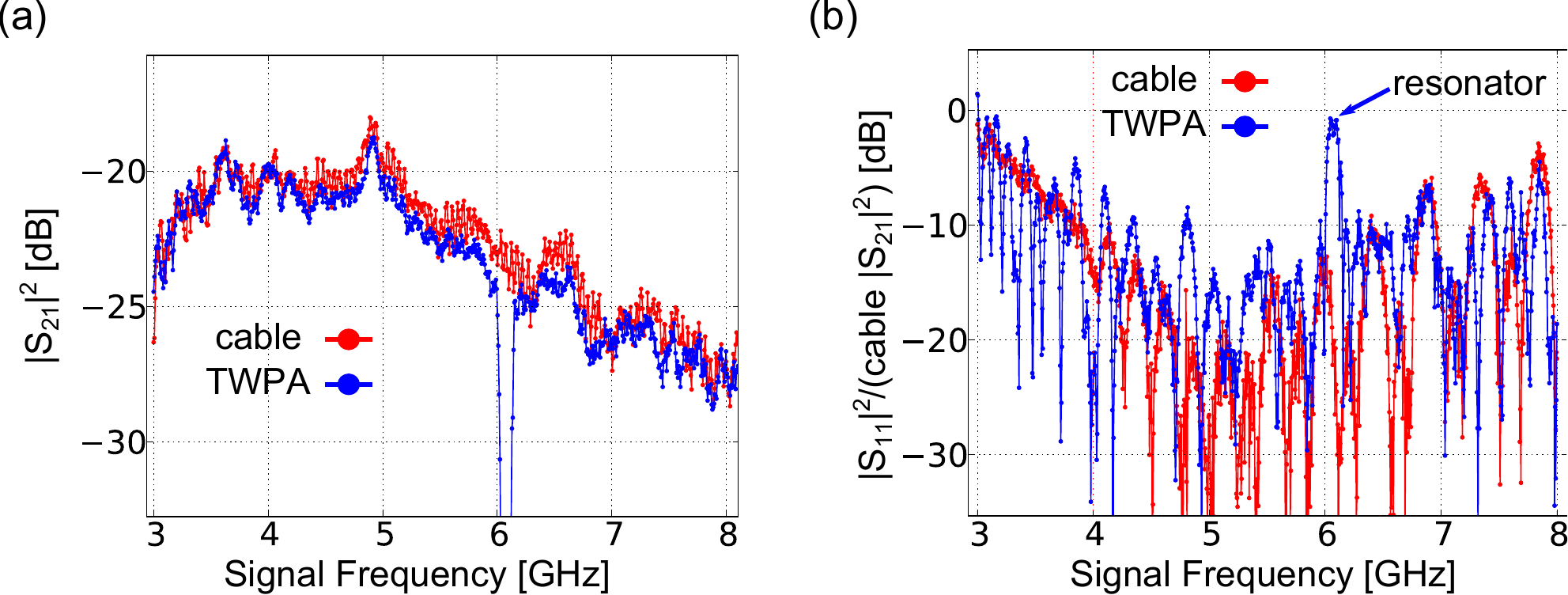}
\caption{(a) Measured transmission amplitude of both a low loss microwave cable (red) and the TWPA (blue).  The TWPA shows a slight decrease in transmission over the range but the deviation is only 0.5\,dB. (b)  Measurement of reflection amplitude of the cable (red) and the TWPA (blue) scaled relative to the cable transmission from (a).  The TWPA's higher reflection amplitude is consistent in magnitude with the lower transmission from part (a), which indicates the majority of the difference comes from reflection in the TWPA rather than loss in the materials.}
\label{fig:transmission}
\end{figure}

The data from this experiment is shown in Fig.\,\ref{fig:transmission}. Figure \,\ref{fig:transmission} (a) shows that the TWPA and cable have the same transmission profile over the majority of the bandwidth.  There is an average decrease in transmission of 0.5\,dB relative to a copper cable, except at the resonator frequency (6.1\,GHz) where there is a significant amplitude dip.  The $\text{S}_{21}$ data for both devices is shown in Fig.\,\ref{fig:transmission} (b) with both data sets scaled relative to $\text{S}_{21}$ for the cable.  The TWPA reflection is in general less than -10\,dB relative to the cable transmission, which is consistent with -0.5\,dB less transmission.  These two data sets taken together suggest that any drop in transmission through the TWPA comes from reflections rather than loss in the transmission line.

\subsection{Characterizing Noise temperature}

\begin{figure*}
  \centering
    \includegraphics[width=1.\textwidth]{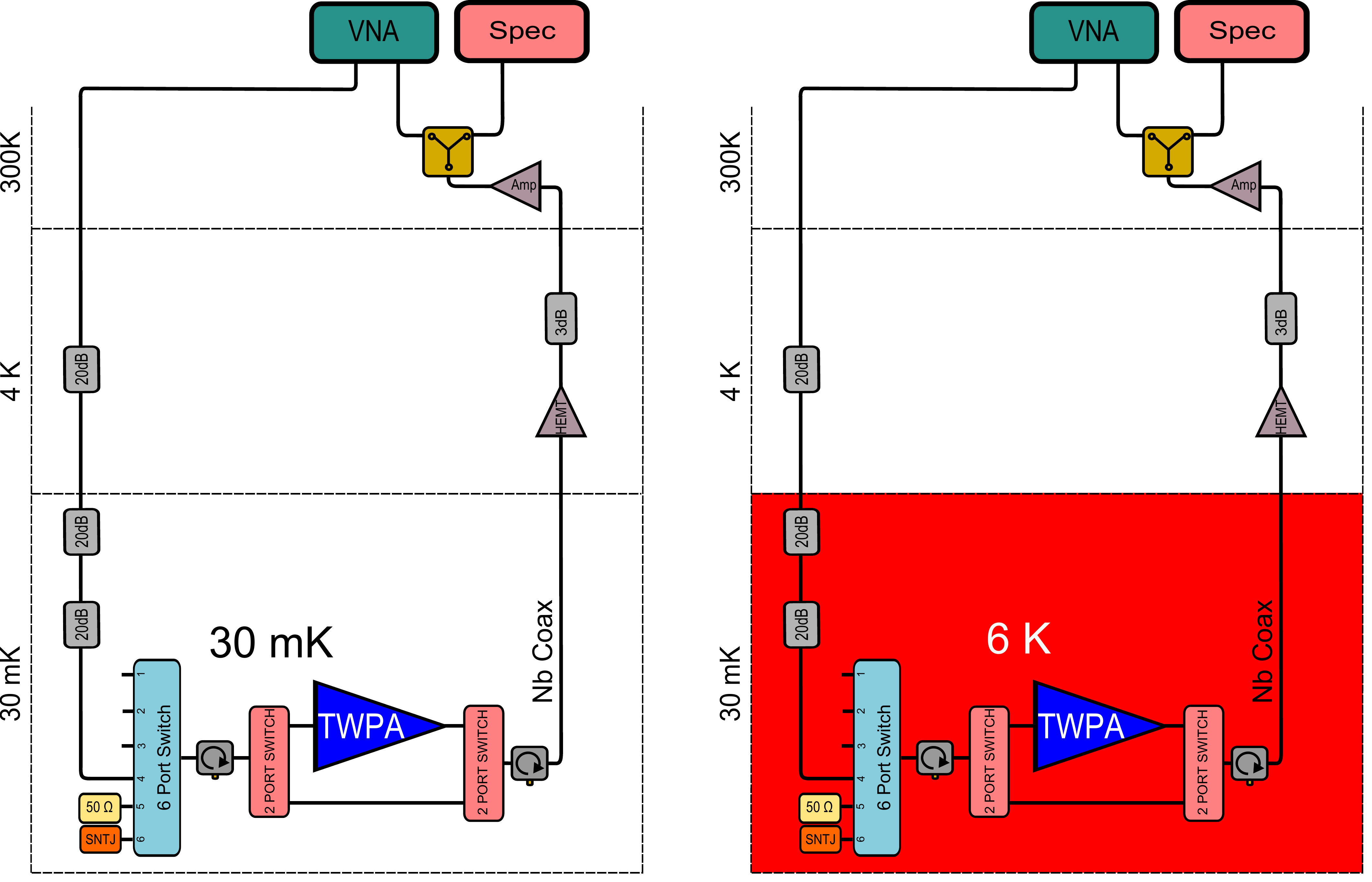}
  \caption{Schematic for paramp experimental setup used to characterize noise.  The left figure shows the experimental setup with paramp, circulators, and microwave switches at base temperature.  The 50\,$\Omega$ on the 6 port switch is heated to calibrate the HEMT noise.  The right schematic shows the portion of the fridge heated (red) to perform a y-factor measurement.  When this calibration is done the 2 port switches are are set to the straight through path which provides 2 circulator channels between the 50\,$\Omega$ and the HEMT, in constrast to the TWPA which is separated from the HEMT by just one circulator.  As circulators are the dominant source of loss, the HEMT system noise seen by the paramp and the 50\,$\Omega$ should only differ by 0.5-1 dB.  This however would skew the system noise higher so it should at least provide an upper bound to the TWPA added noise.}
  \label{fig:yFactor}
\end{figure*}

The TWPA system noise values displayed in this paper were calculated using the method of signal to noise ratio improvement \cite{hatridge:LJPA,JPADesign} discussed in the main text.  In this method the amplified noise and transmission amplitude is first measured when the amplifier is turned off.  The amplifier is then turned on and the amplified noise and transmission amplitude are once again measured.  By comparing the increase in transmission power (gain) to the increase in amplified noise we can measure system noise amplified by the TWPA, provided we know the system noise amplified by the HEMT.  In this measurement signal loss between the TWPA and the HEMT can make this ratio seem smaller and must be taken into account when measuring the amplified HEMT noise. In our setup, shown in Fig. \ref{fig:yFactor}, we use a y-factor measurement \cite{pozar} with a heated 50\,$\Omega$ NiCr resistor on the cold plate of our refrigerator.  In this setup both the TWPA is switched out of the line leaving a 50\,$\Omega$ resistor connected to the HEMT by low loss copper microwave flex cables at 30\,mK, 1-2 circulators, and a Nb coaxial cable connected between 30\,mK and the HEMT at 4\,K.  Due to the difficulty of heating just the 50\,$\Omega$ resistor we use a method in which the entire cold plate of the refrigerator in heated to a much larger temperature (6\,K) and allowed to stabilize before a measurement is performed.  The HEMT amplifier is on a different plate and its temperature is held steady over the course of this experiment.  This methodology, while allowing for accurate temperature measurement of the resistor, can mis-characterize the effect of loss between the resistor and the HEMT.  Any dissipative loss coming from attenuation on the 30\,mK plate would add noise to the signal as it was also at the higher 6\,K temperature.  We assume the dominant source of potential loss comes from the microwave circulators, as the superconducting and copper cables should have negligible loss at these temperatures.  The circulator insertion loss was measured at room temperature to be between 0.5 and 1\,dB.  To account for this we have added 1\,dB error bars to our measurement of the HEMT noise which are in turned scaled to give error bars for the signal to noise ratio improvement of the TWPA.  A second complication arises from the small gain of the TWPA in this experiment.  We were unable to achieve the high 12-14 dB gain displayed earlier in the noise characterization experiment.  Rather the SNR improvement was measured with an average gain of 6-8 dB.  At this level the TWPA gain is not enough to overcome the added noise of the HEMT at 4K.  Thus the lowest system noise we achieve is between 600-700 mK or 2 photons.  After accounting for the residual HEMT noise these numbers are consistent with a single photon of amplified quantum noise.
